\PassOptionsToPackage{table,usenames,dvipsnames}{xcolor}

\documentclass[
]{ceurart}

\usepackage[utf8]{inputenc}
\usepackage{float}
\usepackage{multirow}
\usepackage{multicol} 
\usepackage{array}
\usepackage{subcaption}

\usepackage[labelfont=bf]{caption}
\usepackage{booktabs} 
\usepackage{graphicx} 
\usepackage{caption}   
\usepackage{tabularx} 
\usepackage{adjustbox} 

\definecolor{gray2}{rgb}{0.95, 0.95, 0.95}

\usepackage{listings}
\begin{document}

\copyrightyear{2025}
\copyrightclause{Copyright for this paper by its authors.
  Use permitted under Creative Commons License Attribution 4.0
  International (CC BY 4.0).}

\conference{CSEDM'25: 9th Educational Data Mining in Computer Science Education (CSEDM) Workshop,
  July 20, 2025, Palermo, Sicily, Italy}

\title{Investigating the Impact and Student Perceptions of Guided Parsons Problems for Learning Logic with Subgoals}


\author{Sutapa Dey Tithi}[
orcid=0009-0007-5815-882X,
email=stithi@ncsu.edu
]
\cormark[1]

\author{Xiaoyi Tian}[%
email=xtian9@ncsu.edu,
]

\author{Min Chi}[%
email=mchi@ncsu.edu,
]

\author{Tiffany Barnes}[%
email=tmbarnes@ncsu.edu,
]

\cortext[1]{First Author (led and carried out most of the work)}

\begin{abstract}
Parsons problems (PPs) have shown promise in structured problem solving by providing scaffolding that decomposes the problem and requires learners to reconstruct the solution. However, some students face difficulties when first learning with PPs or solving more complex Parsons problems. This study introduces Guided Parsons problems (GPPs) designed to provide step-specific hints and improve learning outcomes in an intelligent logic tutor. In a controlled experiment with 76 participants, GPP students achieved significantly higher accuracy of rule application in both level-end tests and post-tests, with the strongest gains among students with lower prior knowledge. GPP students initially spent more time in training (1.52 vs. 0.81 hours) but required less time for post-tests, indicating improved problem solving efficiency. Our thematic analysis of GPP student self-explanations revealed task decomposition, better rule understanding, and reduced difficulty as key themes, while some students felt the structured nature of GPPs restricted their own way of reasoning. These findings reinforce that GPPs can effectively combine the benefits of worked examples and problem solving practice, but could be further improved by individual adaptation.

\end{abstract}


\maketitle

\section{Introduction}
Parsons problems have emerged as a promising scaffold for teaching structured problem solving in logic education, which enables learners to reconstruct jumbled proof steps into valid solutions while reducing cognitive load \cite{prather2022scaffolding}. While Parsons problems broken into smaller sub-problems or subgoals can effectively train problem solving skills at a low difficulty level, they often pose challenges when students first encounter them, or the proof structure is complex. The open-ended nature of Parsons problems can make it difficult for these learners to determine which rules to apply and how to connect logical steps into well-structured arguments \cite{shabrina2023learning}. These challenges reflect broader limitations in teaching problem solving techniques. Worked examples often lead to passive engagement by not clearly explaining the reasoning behind each step \cite{renkl2000studying}. This can make it difficult for students to understand why certain choices were made. Conversely, unstructured problem solving can place high cognitive demands on students as they try to construct multi-step proofs \cite{sweller1988cognitive}.

To address these gaps, we introduce \textit{Guided Parsons problems (GPPs)}, a new problem solving approach to augment subgoal-oriented Parsons Problems by providing step-specific hints in an intelligent logic tutor. In GPP, proofs are divided into chunks or ``subgoals''. These subgoals are data-driven groupings of logic statements that represent key logical units{\textemdash}while embedding contextual hints that clarify relevant rule application (e.g., \textit{``Apply Simplification here to isolate ¬P.''}). We also designed a self-explanation module to understand students' perceptions of GPPs. After each GPP, students described how the GPP subgoals helped them. GPPs were designed to promote more active student engagement in problem solving during the guided practice of partially-worked examples.

We deployed our tutor with GPPs in an undergraduate classroom of CS majors, and conducted a controlled experiment. In the controlled experiment, we implemented two training conditions: 1) the Control group who received worked example (WE) and problem solving (PS) logic-proof construction problems, and 2) the GPP group who received Guided Parsons problems (GPPs) along with PS. We used a mixed-methods analysis approach to analyze the impact of GPPs on learning outcomes quantitatively and student perceptions qualitatively. In this study, we investigate the following research questions:

\begin{itemize} 
    \item \textbf{\textit{RQ1}}: What is the impact of Guided Parsons problems (\textbf{GPP}) on student performance and learning outcome?  
    \item \textbf{\textit{RQ2}}: To what extent does student proficiency level moderate the relationship between \textbf{GPPs} and student learning outcomes?
    \item \textbf{\textit{RQ3}}: What common themes emerge from students' self-explanations on their learning experiences with \textbf{GPPs}?
\end{itemize}

\section{Background}\label{sec:lit_review}
GPPs build on the principles of Cognitive Load Theory \cite{sweller1988cognitive} and scaffolded problem solving frameworks \cite{wood1976role}. Sweller et al. \cite{sweller1988cognitive} suggested three types of cognitive load: intrinsic (inherent to the material and may vary based on prior knowledge), extraneous (unnecessary processing, may vary based on how information is presented), and germane (productive mental effort). They argued that learning outcomes are optimized when the intrinsic load is managed, the extraneous load is minimized, and the germane load is promoted.  

Worked examples have been shown to reduce the intrinsic cognitive load and improve learning \cite{paas2003cognitive}. However, Nievelstein et al. found that worked examples may not be beneficial for students with high prior knowledge when problems are structured \cite{nievelstein2013worked}. On the other hand, Renkl et al. demonstrated that worked examples are most effective when they provide instructional explanations or rationales for the solution steps \cite{renkl2005worked}.

Parsons problems, which can be considered as partially worked examples, require students to construct a solution from a given set of jumbled solution steps \cite{denny2008evaluating}. In programming education, Parsons problems have been extensively explored and found to improve students' code writing abilities \cite{weinman2021improving,karavirta2012mobile,denny2008evaluating,ericson2018evaluating}. Poulsen et al. showed that Parsons problems reduced the difficulty in constructing mathematical proofs \cite{poulsen2022evaluating}. Understanding high-level contextual significance \cite{prather2022scaffolding} and subgoal labels \cite{morrison2016subgoals} can help students solve Parsons problems and improve their learning outcomes. Shabrina et al. demonstrated that data-driven, subgoal-oriented Parsons problems can enhance students' subgoaling skills in solving propositional logic proofs \cite{shabrina2023learning}. However, they also found that students struggle with Parsons problems when they first encounter this type of structured or chunked problem or when the connections among different parts of the problem are complex. These results suggest that the design of Parsons problems and their support have important implications for learning.

In this study, we explore GPPs, a new graphical representation of Parsons problems with step-specific hints attached, to improve students' problem solving skills in the context of logic-proof problems. GPPs decompose the proof structure into chunks or subgoals that group statements into logically meaningful units, aligning with Renkl's concept of ``meaningful building blocks'' \cite{renkl2000studying}.  Renkle et al. also emphasized the importance of reflecting on the general aspects of specific problem solutions for transfer learning. Margulieux et al. showed that learners' explanations of the problem solving process can lead to better problem solving performance \cite{margulieux2017using}. To understand the impact of students' self-explaining problem subgoals, we incorporate a self-explanation module in GPPs. Additionally, we integrate step-specific hints into GPPs to address the ``rationale gap'' identified in traditional worked examples \cite{renkl2002worked}. Thus, our intervention is designed to maintain low intrinsic load through subgoals and step-specific scaffolding while facilitating active problem solving participation, a balance that research suggests optimizes cognitive engagement \cite{koedinger2012knowledge}.

\section{Tutor Context}
\begin{figure}[t]
    \centering
    
    \begin{subfigure}[b]{0.45\textwidth}
        \centering
        \includegraphics[width=\textwidth]{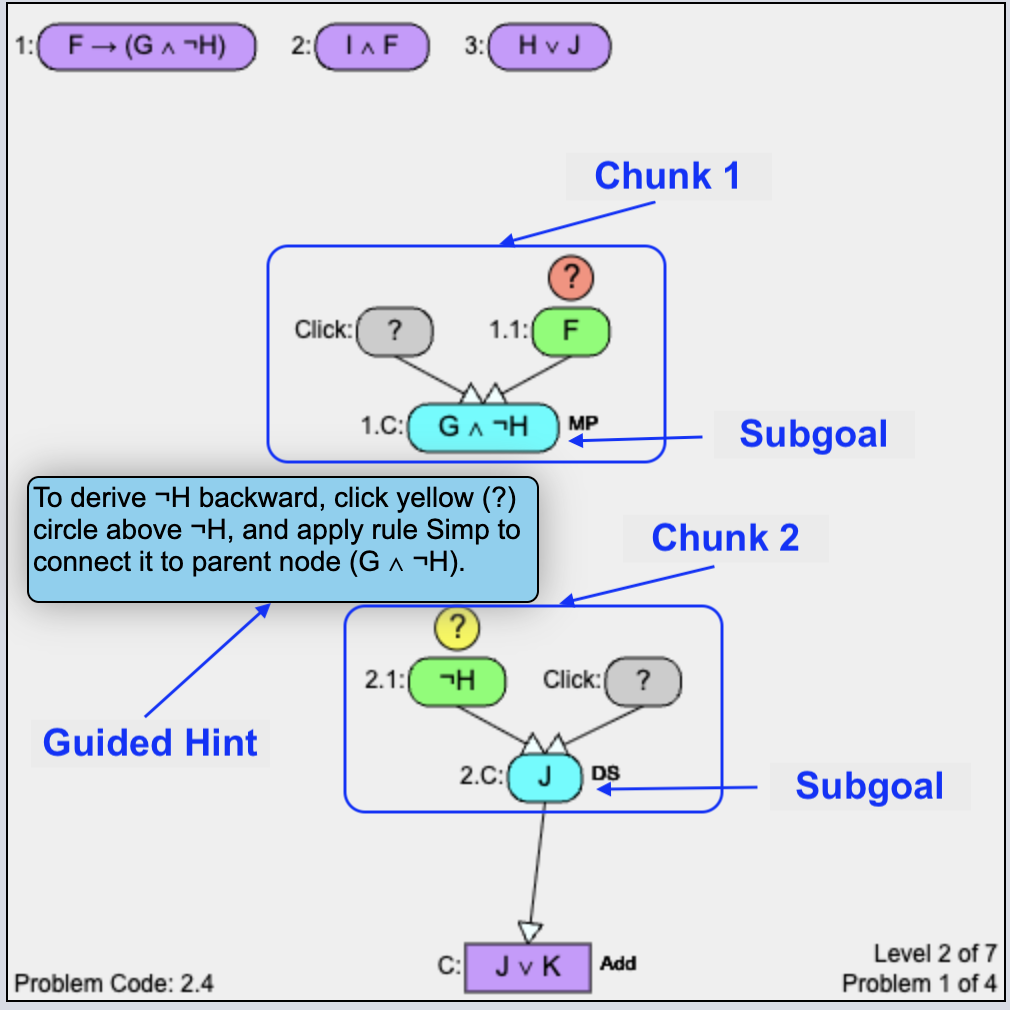}
        \caption{GPP: Guided Parsons Problem}
        \label{fig:gpp}
    \end{subfigure}
    \begin{subfigure}[b]{0.417\textwidth}
        \centering
        \includegraphics[width=\textwidth]{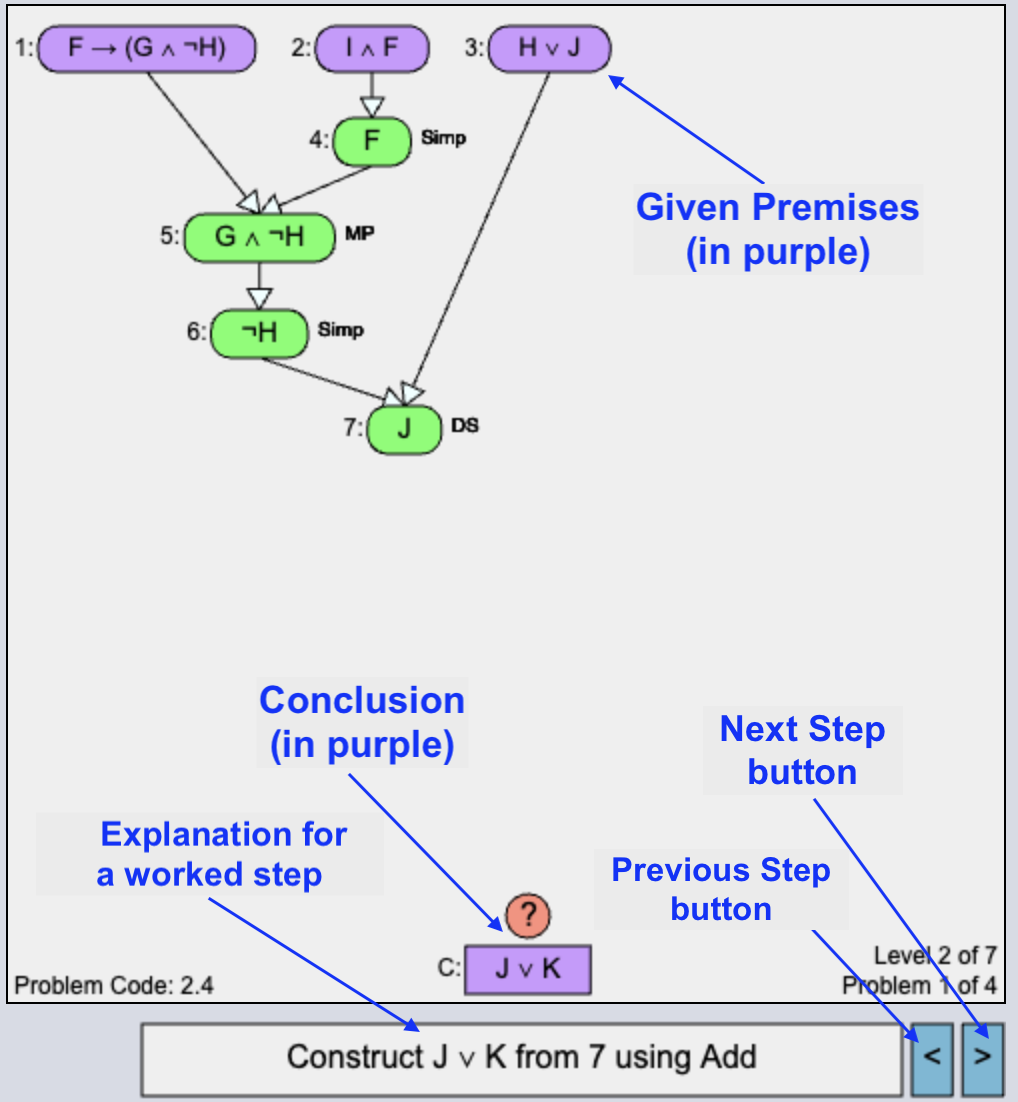}
        \caption{WE: Worked Example}
        \label{fig:we}
    \end{subfigure}
    \hspace{0.03\textwidth}
    \begin{subfigure}[b]{0.4\textwidth}
        \centering
        \includegraphics[width=\textwidth]{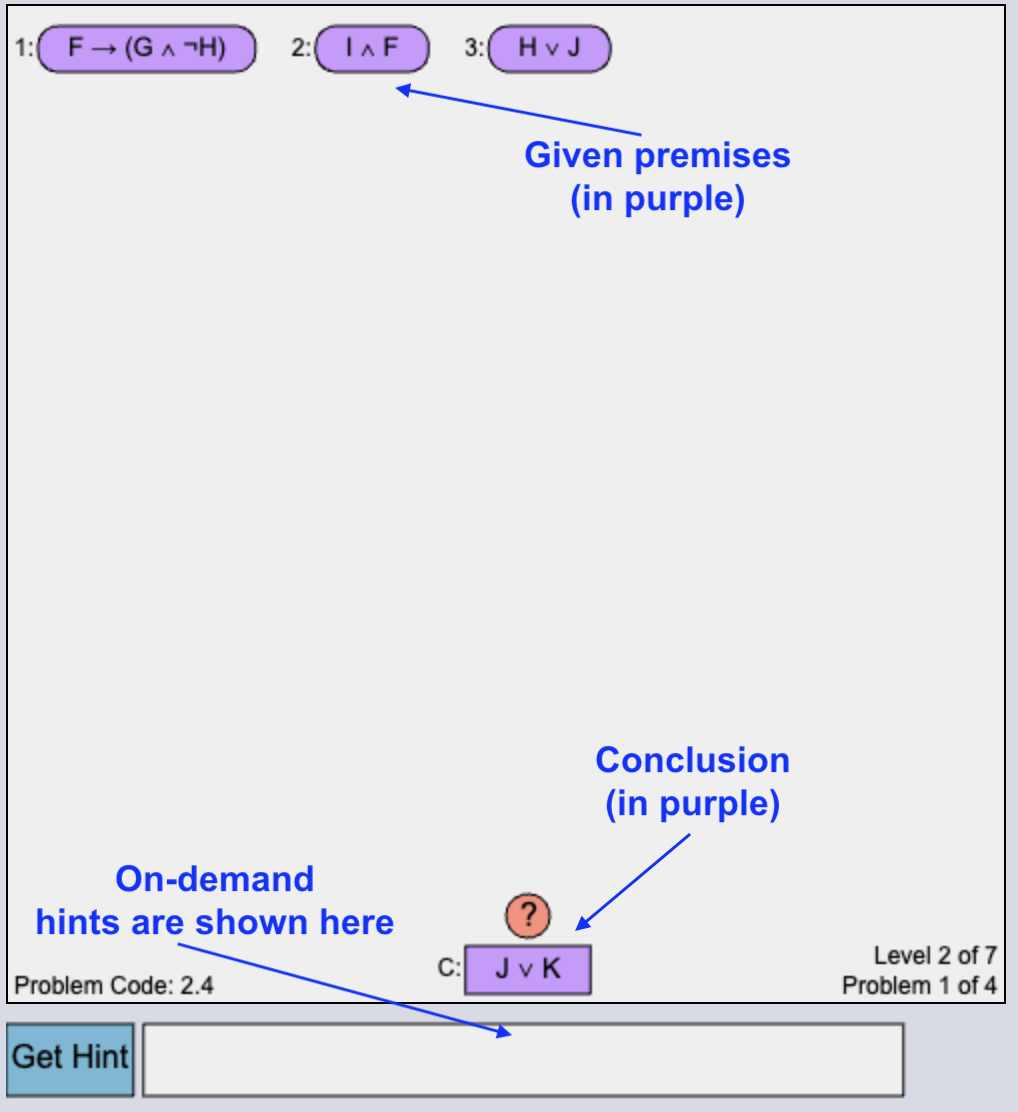}
        \caption{PS: Problem Solving}
        \label{fig:ps}
    \end{subfigure}

    \caption{The tutor interfaces for three different problem types: (a) Guided Parsons Problems (GPP), where some steps are already done and hints are provided for students to complete missing steps, (b) Worked Example (WE), where the tutor performs each step for students, and (c) Problem Solving (PS), where students derive all the steps (nodes) of a proof.}
    \label{fig:problem_types}
\end{figure}

Our logic tutor teaches how to construct propositional logic proofs where a set of given premises and a conclusion are presented as visual nodes for each problem in a graphical representation, as shown in Figure \ref{fig:problem_types}. Students iteratively derive new logic statement nodes to complete each proof. A new logic statement node can be derived working forwards by clicking on 1-2 parent nodes and a logic rule, or working backwards by clicking on a node and hypothesizing a rule and parent nodes to justify it.

We measure students' prior competency based on how they solve two pretest problems. Next, the training session consists of five ordered levels of increasing difficulty, and each level consists of four problems. The last problem in each training level is a \textit{level-end test} problem to measure how the student learns at that level. The posttest level (level 7) consists of six problems. During the pretest, training level-end test, and posttest problems, the tutor offers no help. For each problem, the students receive a score between 0 and 100 based on efficient proof construction, with higher scores corresponding to attempts with smaller solution size, higher accuracy of rule application, and shorter time.

Based on the intervention, the tutor presents three types of problems during training: \textbf{worked examples (\textit{WEs}), problem solving (\textit{PS}), and guided Parsons problems (\textit{GPPs}}).

Worked examples (WEs) are solved by the tutor step-by-step as the students click on the next step \textbf{(>)} button (Figure \ref{fig:we}). On the other hand, PS problems require students to derive all the steps (nodes) of a proof and provide a justification for each step using logic rules (edges) applied to parent logic statement nodes (Figure \ref{fig:ps}). Along with WE and PS problem types, our intervention presents a new problem type, Guided Parsons problem (GPP), where a partially solved proof is presented, and students must complete it by justifying a few of the nodes by selecting a rule and parent nodes to derive them. 

A clustering algorithm was used on previous years' student interaction log data to extract the most commonly related steps as intermediate goals/subgoals, and these subgoals were verified by an expert instructor to be important in solving each tutor problem \cite{shabrina2023learning}. The most commonly related steps are presented together as a chunk on the screen. Figure \ref{fig:gpp} shows a screenshot of the initial presentation of a guided Parsons problem (GPP). At the top of the screen, the problem’s given statement nodes 1, 2, and 3 are shown in purple. The goal is to derive the conclusion (the purple node labeled with C: $J \lor K$), and the complete proof is a connected graph with edges from the givens to the conclusion. Generally, to solve a logic proof in DT, students derive new nodes by clicking on given statement nodes and a rule, until the conclusion is derived. A derived statement node is said to be justified when it has arrows from its parent nodes to the derived node, labeled with the rule that justifies the statement. For example, in Figure \ref{fig:gpp}, the node 2.C for statement J is a parent node with the Addition (Add) rule to derive the conclusion C: $J \lor K$, illustrated with an arrow from node 2.C to the C, the conclusion. 
Each GPP provides students with all the statement nodes needed to complete a proof, but students must add a few justifications to connect all the nodes to one another with missing edges for rules. The nodes without incoming edges are unjustified. GPPs guide students to justify each unjustified node by specifying the rule used to derive it. In Figure \ref{fig:gpp}, statement 2.1: $-H$  can be derived from 1.C: $G \land -H$ using Simplification. GPPs guide students with a hint, as shown in the Figure \ref{fig:gpp}, to derive the statement. To complete this step, students click on the yellow question mark above 2.1, choose the rule Simp, and click on statement 1.C to show that there should be an edge from 1.C to 2.1. GPPs are divided into chunks, where important subgoals in the problem are shown in light blue/cyan, grouped with the nodes used to derive them. For example, in this problem, there are 2 chunks 1 and 2, with subgoals 1.C and 2.C respectively, that are needed to complete the problem. DT guides students using popup hints with instructions to work backwards from the conclusion to connect to node 2.C, then connect 2.1 to chunk 1’s conclusion 1.C, and then 1.1 to the givens. 

\section{Methods} \label{sec:data-collection}
\textbf{Experimental Conditions. }We designed two training conditions in the logic tutor:
\begin{itemize}
    \item \textbf{{Control:}} Students assigned to the \textit{Control} condition received PS or WE (selected randomly) during training.
    \item \textbf{{GPP:}} Students assigned to \textit{GPP} condition received PS or GPP (selected randomly) during training.
\end{itemize}

The tutor was deployed with the two training conditions in an undergraduate Discrete Mathematics course at a public research university in the United States in the Spring of 2024. We did not collect course-specific demographics; it is noteworthy that discrete math is a mandatory course for all CS majors. Therefore, for an approximation, we report the demographics of the 2021-22 graduating class of CS majors with the gender composition of 83\% men and 17\% women; and race/ethnicity of 58\% white, 18.5\% Asian, 3\% Hispanic/Latin, 2\% Black/African American, 9\% other races, with the remaining 9.5\% having international student status for whom race/ethnicity information was not available. This study was approved by the university IRB, and only authorized researchers could access the data collected from the participants. 

Each participating student in that course was assigned to one of the two training conditions after they completed the pretest problems. We compare only students who completed all 7 tutor levels, with 30 students in the Control group and 46 students in the GPP group. 

\textbf{Data collection and analysis. }
For both training conditions, we analyzed their \textbf{interaction logs} to measure their learning and performance. For students in the GPP group, we collected \textbf{students' self-explanation responses} after solving each GPP problem (e.g., \textit{``How did the subgoals $(G \land ¬H), J$ help you derive the conclusion?''}). Research shows that self-explanation promotes learning \cite{vanlehn1992model,bisra2018inducing}.
Answering the self-explanation question was mandatory, and a total of 326 explanations were collected from GPP students. We hoped that self-explanations would help students use the GPPs to better understand the structure of logic proofs and how a directed strategy guides experts to build subgoals that link the givens to the conclusions. Therefore, we conducted a thematic analysis of the self-explanations to determine whether and how students were learning about subgoals through GPPs (RQ3). The themes were derived through an inductive coding process \cite{thomas2006general} following established thematic analysis methodology \cite{guest2011applied,muller2014curiosity}. Two researchers independently coded a subset of explanations (initial agreement: 91.9\%), and then they discussed any code discrepancies. After reaching consensus on the codes, one researcher coded the remaining self-explanations. 

\textbf{Performance Metrics.} A student’s problem score is a combination of normalized metrics for the \textbf{problem completion time}, \textbf{total number of steps}, and \textbf{rule application accuracy} on a single problem, which ranks a student based on how fast, efficient, and accurate they are. Total steps in a problem include any attempt a student makes to derive a new node (including mistakes). Rule application accuracy is the total number of correct rule applications divided by all rule application attempts.

\section{Results}
We report the results by research questions RQ1 on student learning, RQ2 on GPP effectiveness, and RQ3 on GPP self-explanations.
\subsection{RQ1: Student Performance and Learning}
To evaluate the impact of GPPs on learning outcomes, we analyzed students’ performance scores and normalized learning gains (NLG) across training conditions. Our analyses focused on performance metrics derived from the training level-end test problems (2.8, 3.8, 4.8, 5.8, and 6.8) and the posttest problems (7.1–7.6). Students solved these problems independently without any tutor help. We performed a combination of statistical comparisons, including mixed-effects regression and Mann-Whitney U tests, to account for the non-normal nature of our data and problem-specific variability.

\begin{table}[t]
\renewcommand{\arraystretch}{1.5}  
\caption{Overall problem score and normalized learning gain (NLG) (Mean (SD)) across the two training groups. [Note: \textbf{Boldface} indicates comparatively better averages]}
\label{tab:NLG}
\setlength{\tabcolsep}{2pt}
\centering
\small
\begin{tabular}{l|l|l|l|l}

\hline
\cellcolor{gray2}\begin{tabular}[c]{@{}l@{}}\textbf{Group} \\ \textbf{(N)}\end{tabular} & \cellcolor{gray2}\begin{tabular}[c]{@{}l@{}l@{}}\textbf{Pre}\end{tabular}          & \cellcolor{gray2}\begin{tabular}[c]{@{}l@{}l@{}}\textbf{Post}\end{tabular}         & \cellcolor{gray2}\begin{tabular}[c]{@{}l@{}}\textbf{NLG} \end{tabular}         &  \cellcolor{gray2}\begin{tabular}[c]{@{}l@{}} \textbf{\%Students} \\ \textbf{with} \\ \textbf{(+) NLG} \end{tabular} \\ \hline
Control (30)                                              & 62.8 (18.7) & 70.4 (14.4) & 0.26 (0.45) & 73\% \\ \hline
GPP (46)                                               & 63.8 (18.1) & \textbf{72.6 (8.2)} & \textbf{0.27 (0.44)} & \textbf{78\%} \\ \hline
\end{tabular}

\end{table}



\textbf{Problem Score \& NLG.}
In the pretest problems, there were no significant differences in problem scores across the two conditions (Control = 62.8, GPP = 63.8, p > 0.05). We conducted mixed-effect regression analyses to examine the relationship between training conditions and problem score, with problem IDs as random effects and training conditions as fixed effects. For training level-end test problems, results indicated a marginally significant association ($\textbf{p = 0.055}$) between training conditions and problem score (Control = 57.9, GPP = 62.8). Although problem scores were not significantly different in posttest problems across conditions ($p = 0.46$) (Control = 70.4, GPP = 72.6), GPP students demonstrated better average scores in posttest problems (Table \ref{tab:NLG}).

To identify the effectiveness of the training conditions in promoting learning, we analyzed students’ normalized learning gain (NLG) across the two training conditions. Normalized learning gain is calculated using the average problem scores on the pre and posttest problems  with the equation \cite{shabrina2023learning}:
\begin{equation} \text{NLG} = \frac{(\text{posttest score}-\text{pretest score})}{\sqrt{(100 - \text{pretest score})}}
\end{equation}

Note that, we normalize NLG scores between -1 and 1. In our analyses, NLG may show negative numbers. These negative numbers are due to the difficulty of the pretest section compared to the much higher difficulty of posttest section and does not indicate negative learning. Table \ref{tab:NLG} shows there was no significant difference in NLG between the two conditions, although the GPP group yielded higher positive NLG rates (78\% in GPP vs. 73\% in Control), moving their NLG toward positive more often.

\textbf{Time.}
The training time for GPP was significantly higher than the control group (GPP: 1.52 hrs vs. Control: 0.81 hrs). But in the training level-end test problems as well as posttest problems, GPP students spent less time on average (no significant difference), suggesting improved problem solving efficiency (Table \ref{tab:metrics_comparison}).
\begin{table}[ht]
    \centering
    \caption{Time comparison across Control and GPP Groups. All values are written in hours (Mean (SD)). [Note: \textbf{\textcolor{blue}{Blue\textsuperscript{*}}} indicates a significant difference.]}
    \label{tab:metrics_comparison}
    \begin{tabular}{@{}lcc@{}}
        \toprule
        \textbf{Time}              & \textbf{Control} & \textbf{GPP} \\ \midrule
        Training                 & \textbf{\textcolor{blue}{0.81 (0.37)\textsuperscript{*}}}                  & 1.52 (0.77)             \\
        Level-End test       & 1.87 (1.43)                & 1.50 (0.92)             \\
        Posttest                 & 1.02 (0.96)                & 0.91 (0.59)             \\
        Total Tutor             & 4.71 (2.31)                & 4.75 (1.76)             \\ 
        \bottomrule
    \end{tabular}
\end{table}


\begin{figure}[t]
    \centering

    \begin{subfigure}[b]{0.48\textwidth}
        \centering
        \includegraphics[width=\linewidth]{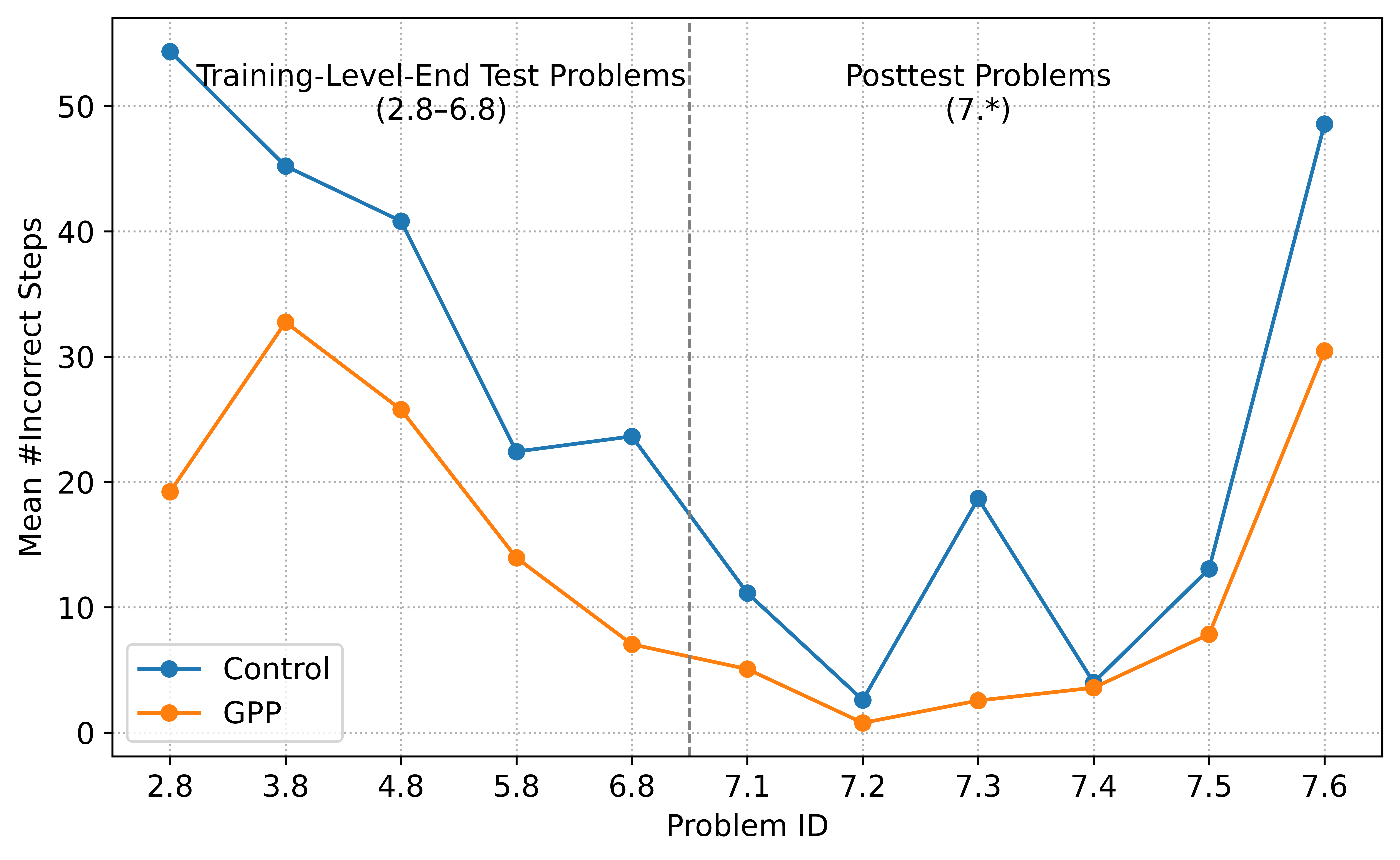}
        \caption{Average number of incorrect steps across conditions in the training level-end test and posttest problems.}
        \label{fig:mean_incorrect}
    \end{subfigure}
    \hfill
    \begin{subfigure}[b]{0.48\textwidth}
        \centering
        \includegraphics[width=\linewidth]{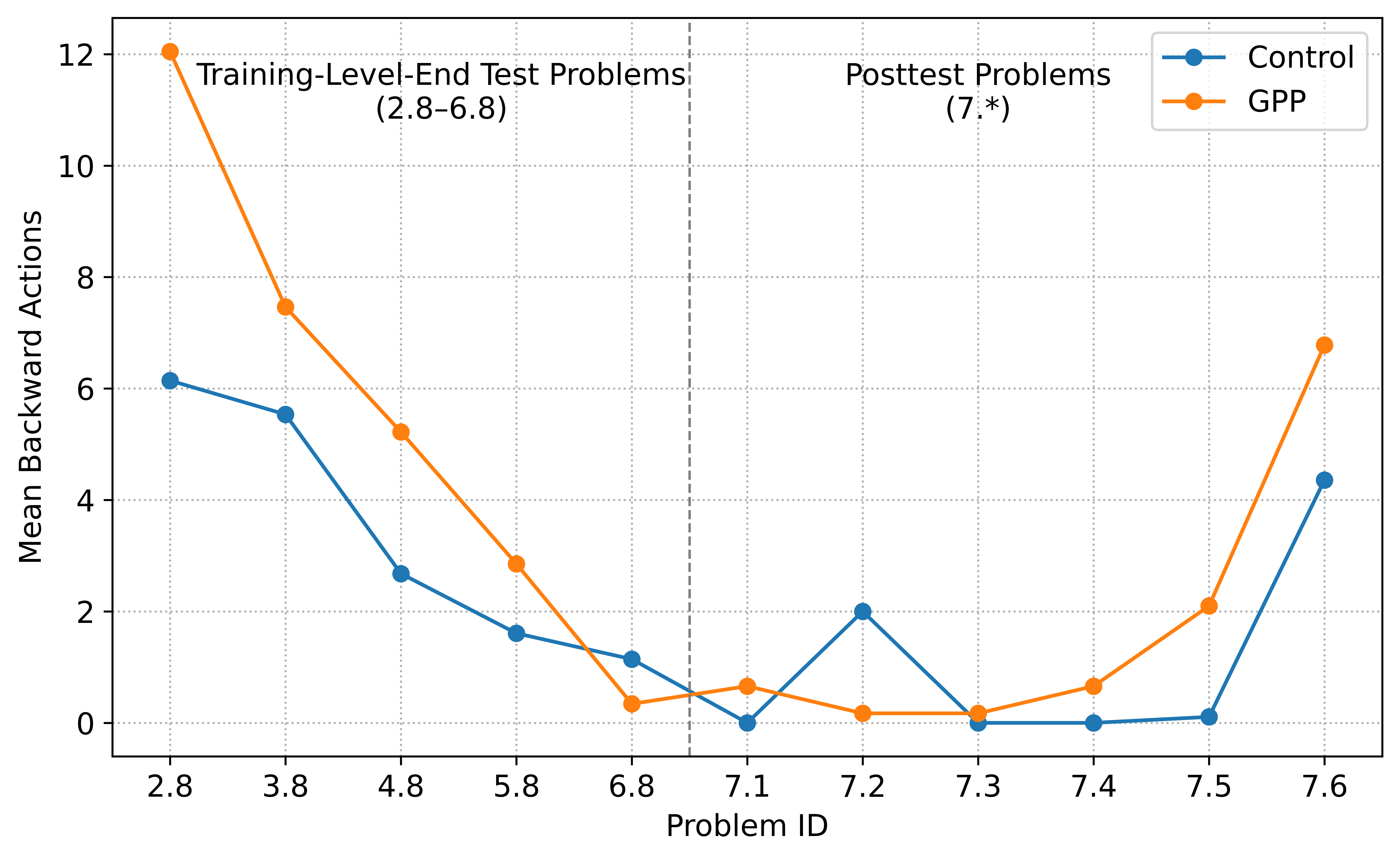}
        \caption{Average number of backward attempts across conditions in the training level-end test and posttest problems.}
        \label{fig:backward}
    \end{subfigure}

    \caption{Comparisons of incorrect steps and backward attempts across conditions during training level-end test and posttest phases.}
    \label{fig:wrong_backward_comparison}
\end{figure}

\textbf{Rule Application Accuracy.} In addition to the overall problem score, we observed students' step derivation behavior. Total Correct Steps are the number of correct rule applications over the whole tutor. Total Incorrect Steps are the number of rule applications that are either incorrect by virtue of selecting a rule that does not apply to the arguments (e.g. using the rule \textit{Simplifaction} on a logical statement that cannot be simplfied) or not being able to correctly identify what statement a rule application would derive. There was a significant difference in the average number of incorrrect steps between groups in the level-end tests (Control = 37.4, GPP = 19.6, $\textbf{p < 0.001}$) and a marginal difference in incorrect steps in the final posttest (Control = 16.4, GPP = 8.4, $\textbf{p = 0.06}$). The trend in the average number of incorrect steps is shown in Figure \ref{fig:mean_incorrect}. 

Rule application accuracy is defined as the number of total correct rule applications divided by the total number of attempts for rule application. Mann-Whitney U tests indicate that, training with GPP problems significantly improved the rule application accuracy among the students in both training level-end test problems, and the posttest problems (Table \ref{tab:mann_whitney_results}).

\begin{table}[t]
    \centering
    \small
    \caption{Rule accuracy (Mean (SD)) across two conditions in training level-end test and posttest problems. [Note: \textbf{\textcolor{blue}{Blue\textsuperscript{*}}} indicates a significant difference.]}
    \label{tab:mann_whitney_results}
    \begin{tabular}{@{}lccc@{}}
        \toprule
        \textbf{Test}          & \textbf{Control } & \textbf{GPP} & \textbf{Test Results}        \\ \midrule
        Level-End Tests         & 59.4 (24.6)                    & \textbf{\textcolor{blue}{68.7 (20.2)\textsuperscript{*}}}                    &  $p$ = .002 \\ \midrule
        Posttests              & 72.7 (22.1)                    & \textbf{\textcolor{blue}{79.8 (17.2)\textsuperscript{*}}}                    & $p$ = .003 \\ \bottomrule
    \end{tabular}
    
\end{table}

\textbf{Backward Actions.} As a measure of students’ response to the training with backward strategy in GPP, we count their independent attempts to work backwards. In Figure \ref{fig:backward}, we report mean backward attempt counts for the training level-end test and posttest problems. In the posttest problems, GPP students had significantly more backward attempts (Control = 1.1, GPP = 1.8, $\textbf{p = 0.01}$). 

\subsection{RQ2: Impact of GPP by Pretest Score}
Prior research suggests that instructional interventions like worked examples and Parsons problems may have varying effectiveness based on learners' prior knowledge \cite{koedinger2012knowledge,kalyuga2009expertise}. To investigate this phenomenon in our context, we analyzed how the impact of GPPs differed based on prior proficiency. We compared metrics across three phases (pretest, level-end test, posttest) using Mann-Whitney U tests. We used the Bonferroni correction to adjust the significance thresholds. We categorized two prior proficiency groups based on a median split on the pretest score.

As shown in Table~\ref{tab:combined_metrics_highlighted}, GPPs significantly improved rule application accuracy for low prior knowledge students at both level-end test (Control = 51.6, GPP = 68.3, $\textbf{p < .001}$) and posttest (Control = 66.7, GPP = 79.1, $\textbf{p < .001}$) problems. High prior knowledge students showed comparable final rule accuracy across groups (Control = 78.5, GPP = 80.5, $p = 0.68$), potentially suggesting a ceiling effect for learning rules. GPPs reduced redundant steps for high prior knowledge students during level-end test problems (Control = 11.5, GPP = 9.66, $\textbf{p = 0.02}$) but increased steps for low prior knowledge counterparts (Control = 9.71, GPP = 11.64, $\textbf{p = 0.03}$). High prior knowledge students in the GPP condition showed significantly reduced posttest times  ($\textbf{p=0.01}$). Low prior knowledge students showed no significant time differences. These results suggest that GPPs helped advanced learners become more efficient, while they only helped novices gain a better understanding of logic rules.

\begin{table*}[ht]
\centering
\small
\renewcommand{\arraystretch}{1.2} 
\setlength{\tabcolsep}{1pt}
\caption{Performance analysis across two training conditions categorized on pretest scores. [Note: \textbf{\textcolor{blue}{Blue\textsuperscript{*}}} indicates a significant difference. \textbf{Boldface} indicates comparatively better averages (e.g. higher for rule accuracy/lower for extra steps).]}
\label{tab:combined_metrics_highlighted}
\begin{tabular}{@{}llcccccc@{}}
\toprule
\textbf{Metric} & \textbf{Test} & \multicolumn{3}{c}{\textbf{High Prior Knowledge}} & \multicolumn{3}{c}{\textbf{Low Prior Knowledge }} \\ 
\cmidrule(lr){3-5} \cmidrule(lr){6-8}
 & & \textbf{Control (N=15)} & \cellcolor[gray]{0.9}\textbf{GPP (N=23)} & \textbf{p-val} & \textbf{Control (N=15)} & \cellcolor[gray]{0.9}\textbf{GPP (N=23)} & \textbf{p-val} \\
\midrule
\multirow{3}{*}{Rule Accuracy} 
 & Pretest    & 67.8 (28.4) & \cellcolor[gray]{0.9}60.1 (29.2) & 0.22 & 39.9 (20.4) & \cellcolor[gray]{0.9}42.3 (0.20) & 0.23 \\
 & Level-End  & 67.2 (23.0) & \cellcolor[gray]{0.9}\textbf{69.2 (20.6)} & 0.56 & 51.6 (23.5) & \cellcolor[gray]{0.9}\textbf{\textcolor{blue}{68.3 (19.9)\textsuperscript{*}}} & {<0.001} \\
 & Posttest   & 78.5 (21.5) & \cellcolor[gray]{0.9}\textbf{80.5 (16.9)} & 0.68 & 66.7 (21.2) & \cellcolor[gray]{0.9}\textbf{\textcolor{blue}{79.1 (17.4)\textsuperscript{*}}} & {<0.001} \\
 \midrule
\multirow{3}{*}{Step Count} 
 & Pretest    & 5.2 (1.6) & \cellcolor[gray]{0.9}5.0 (1.3) & 0.7 & 6.9 (2.51) & \cellcolor[gray]{0.9}7.3 (3.0) & 0.85 \\
 & Level-End  & 11.1 (4.3) & \cellcolor[gray]{0.9}\textbf{\textcolor{blue}{9.6 (3.6)\textsuperscript{*}}} & {0.02} & \textbf{\textcolor{blue}{9.7 (3.6)\textsuperscript{*}}} & \cellcolor[gray]{0.9}11.6 (5.8) & {0.03} \\
 & Posttest   & \textbf{8.1 (3.2)} & \cellcolor[gray]{0.9}8.2 (3.3) & 0.81 & \textbf{8.6 (3.8)} & \cellcolor[gray]{0.9}9.6 (4.6) & 0.14 \\
\midrule
\multirow{3}{*}{Problem Time (minutes)} 
 & Pretest    & 16.2 (17.2) & \cellcolor[gray]{0.9}12.4 (16.2) & 0.74 & 38.3 (17.4) & \cellcolor[gray]{0.9}31.1 (15.1) & 0.60 \\
 & Level-End  & 19.8 (14.0) & \cellcolor[gray]{0.9}\textbf{18.9 (12.5)} & 0.38 & 24.3 (13.2) & \cellcolor[gray]{0.9}\textbf{20.9 (13.5)} & 0.20 \\
 & Posttest   & 9.3 (8.7)\textsuperscript{*}  & \cellcolor[gray]{0.9}\textbf{\textcolor{blue}{6.2 (9.2)\textsuperscript{*}}} & {0.01} & 13.6 (12.6) & \cellcolor[gray]{0.9}\textbf{10.1 (16.7)} & 0.58 \\

\bottomrule
\end{tabular}
\end{table*}

\subsection{RQ3: GPP Self-explanation Themes} 
To understand student perceptions and experiences with GPPs, we conducted a thematic analysis on 326 unique student explanations from 46 students in GPP group (method detailed in Section \ref{sec:data-collection}) and identified five key themes: \textit{Task Decomposition}, \textit{Rule Understanding}, \textit{Reduced Difficulty}, \textit{Backward Reasoning}, and \textit{Difficulty}. Next, we discuss these emergent themes from student explanations.

\textbf{Task Decomposition} ($N=178$): The most common (mentioned in 54.6\% of explanations) benefit students mentioned about GPPs is that the subgoal-oriented solutions naturally helped them break down the proofs into manageable steps. As one student noted, \textit{``They broke down the problem into more understandable smaller problems that I was able to solve and then piece together.''}, and another student quoted ``\textit{They showed me short-term goals so that I know how to piece together the puzzle pieces.}''. As a result, the GPP made it ``\textit{much less intimidating to solve}'' a logic problem than those without subgoals.

\textbf{Rule Understanding} ($N=124$): Step-specific hints in GPP provided context to explain the connection among different subgoals. Thirty-eight percent of explanations highlighted improved rule understanding, as one student mentioned, \textit{``They helped by giving me a better understanding of the rules needed to complete the problem''} and when and how to apply them appropriately (e.g., \textit{``The hints were useful, I wouldn't really have thought to double negate something like that. It's certainly something I'll keep in mind in future.''}).

\textbf{Reduced Difficulty} ($N=73$): GPP shows a skeleton of the solution first, and thus encourages task planning before working through a problem. In 22\% of explanations, students reported reduced mental cognitive load through guided workflows (e.g., \textit{``Made the problem easier to solve.", ``The problem showed an easy relationship.'', ``It allowed me to work on simpler goals and not get distracted on long mistakes.''}).

\textbf{Backward Reasoning} ($N=31$): GPP hints were designed to help students complete the proof using a backward strategy. Although in our tutor, students derive new statements using forward reasoning more often, GPP could help students practice backward reasoning while being a low-effort training intervention. In 9.5\% of their explanations, students reported that the hints and subgoals helped them perform backward chaining easily on this problem type (e.g., \textit{``They provided obvious stepping stones to move backward through the logic in a readily apparent path.''}).

\textbf{Difficulty} ($N=24$): Some student explanations (7.4\% of explanations) reported struggle with the constrained GPP workflows. For example, one student noted \textit{``It didn't help, and it made the problem harder by disrupting my own way of working through the problem, and forced me to work it the way they want me to.'', ``They made the problem harder, rather than helping.''}. Within this difficulty category, a few explanations reported ``confusion'', implying that parts of the GPP design might confuse students and impose extraneous cognitive load on them.


In summary, student self-explanations seemed to indicate an overall positive experience with GPP problems, with the majority noting the benefits of task decomposition; other benefits include mastery of rule understanding and guided workflow. However, we also noted a potential limitation of GPP, with a small subset of students reporting that the structured nature of GPP might disrupt their approach to independent problem solving. Additionally, we received suggestions for improving the design of GPP. 

\section{Discussion}
Our findings highlight the effectiveness of Guided Parsons problems (GPPs) as an intervention for enhancing problem solving skills. By maintaining a balance between structured scaffolding and student autonomy, GPPs address critical gaps in traditional PPs. The chunking of proofs into semantically meaningful segments, along with step-specific hints, aligns with the principles of Cognitive Load Theory \cite{sweller1988cognitive}, potentially reducing intrinsic cognitive load. This approach proved particularly beneficial for students with low prior knowledge, who demonstrated significant improvements in rule application accuracy. Such improvements also reflect Renkl’s argument that well-structured examples, along with clear rationales, are critical for schema acquisition and deeper understanding \cite{renkl2002worked}.

In addition, moderation analysis reveals that GPP can be helpful in different proficiency levels. While low prior knowledge students benefited primarily from added scaffolding, resulting in fewer incorrect steps, high prior knowledge students benefited from the GPP framework to improve their efficiency, as evidenced by the reduced number of steps (p=0.02). These findings align with the expertise reversal effect \cite{kalyuga2009expertise}, suggesting that learners with high prior knowledge may require less detailed support and can benefit from advanced scaffolding techniques. Future iterations of GPPs could incorporate adaptive hints to ensure that learners with different prior proficiency receive appropriately tailored support.

Student self-explanation on GPPs aligned with the quantitative analysis findings from our RQ1 and RQ2, and highlighted the benefits of having subgoal-oriented proof structures, revealing clear, logical flows, and illustrating the relevance of necessary rules. Three prominent themes, \textit{Task Decomposition}, \textit{Rule Understanding}, and \textit{Reduced Difficulty}, emphasize how the subgoals and step-specific hints made the proofs more manageable, potentially reducing cognitive load. Conversely, several students perceived the structured nature of the proof as disruptive to their own reasoning processes. 
These results suggest that GPPs could be further enhanced by making them adaptive to individual student skill levels, which has been shown to be effective for programming \cite{haynes2022impact}.
This could enable high prior knowledge students to maintain autonomy and exploration, while offering additional scaffolding only as needed.

\section{Conclusion and Future Work}
This study introduces Guided Parsons problems (GPPs) as a novel intervention for logic education, combining the structured guidance of worked examples with active engagement in subgoal-focused problem solving. Our results demonstrate the effectiveness of GPP in improving rule application accuracy and reducing cognitive load{\textemdash}particularly among low prior knowledge students. However, our study had several limitations. The study was conducted in only one ITS platform, limiting generalizability. 
Future research should explore a more adaptive implementation of GPPs to dynamically adjust the amount of scaffolding according to learners’ performance and metacognitive needs. Longitudinal studies could further examine how GPPs influence knowledge retention over extended periods or contribute to transferable problem solving skills.  Future studies may also benefit from measuring the perceived cognitive load using standardized survey questions.

\bibliography{main}

\appendix



\end{document}